\documentclass{vgtc}                % final (journal style)
\ifpdf%                                % if we use pdflatex
  \pdfoutput=1\relax                   % create PDFs from pdfLaTeX
  \usepackage{graphicx}                % allow us to embed graphics files
  \DeclareGraphicsExtensions{.pdf,.png,.jpg,.jpeg} % for pdflatex we expect .pdf, .png, or .jpg files
\else%                                 % else we use pure latex
  \usepackage{graphicx}                % allow us to embed graphics files
  \DeclareGraphicsExtensions{.eps}     % for pure latex we expect eps files
\fi%

\graphicspath{{figures/}{pictures/}{images/}{./}} % where to search for the images

\usepackage{microtype}                 % use micro-typography (slightly more compact, better to read)
\PassOptionsToPackage{warn}{textcomp}  % to address font issues with \textrightarrow
\usepackage{textcomp}                  % use better special symbols
\usepackage{mathptmx}                  % use matching math font
\usepackage{times}                     % we use Times as the main font
\usepackage{cite}                      % needed to automatically sort the references
\usepackage{tabu}                      % only used for the table example
\usepackage{booktabs}                  % only used for the table example
\usepackage{graphicx}
\usepackage{multirow}
\usepackage{balance}
\usepackage{color}
\usepackage[dvipsnames]{xcolor}
\usepackage{xspace}
\usepackage{textcomp}
\usepackage{amsmath}
\usepackage{hyperref}
\usepackage[moderate]{savetrees}
\usepackage{enumitem}

\DeclareRobustCommand{\thetool}{{Foresight}\xspace}

\DeclareRobustCommand{\parhead}[1]{\noindent\textit{#1}} % PJH: formatting command for paragraph heads
\DeclareRobustCommand{\bfparhead}[1]{\noindent\textbf{#1}} % CD: boldface \parhead  

\onlineid{0}

\vgtccategory{Research}

\vgtcinsertpkg

\title{\thetool: Rapid Data Exploration Through Guideposts}

\author{
  \c{C}a\u{g}atay Demiralp\thanks{e-mail: \{cagatay.demiralp, phaas, spartha, tejaswinip\}@us.ibm.com}\\
  \scriptsize IBM Research %
  \and Peter J. Haas\\
  \scriptsize IBM Research %
  \and Srinivasan Parthasarathy\\
  \scriptsize IBM Research %
  \and Tejaswini Pedapati\\
  \scriptsize IBM Research 
}

\abstract{
Current tools for exploratory data analysis (EDA) require users to manually
select data attributes, statistical computations and visual encodings. This can
be daunting for large-scale, complex data. We introduce \thetool, a
visualization recommender system that helps the user rapidly explore large
high-dimensional datasets through ``guideposts.'' A guidepost is a
visualization corresponding to a pronounced instance of a statistical
descriptor of the underlying data, such as a strong linear correlation between
two attributes, high skewness or concentration about the mean of a single
attribute, or a strong clustering of values. For each descriptor, Foresight
initially presents visualizations of the ``strongest'' instances, based on an
appropriate ranking metric. Given these initial guideposts, the user can then
look at ``nearby'' guideposts by issuing ``guidepost queries'' containing
constraints on metric type, metric strength, data attributes, and data values.
Thus, the user can directly explore the network of guideposts, rather than the
overwhelming space of data attributes and visual encodings.  Foresight also
provides for each descriptor a global visualization of ranking-metric values to
both help orient the user and ensure a thorough exploration process. Foresight
facilitates interactive exploration of large datasets using fast, approximate
sketching to compute ranking metrics.  We also contribute insights on EDA
practices of data scientists, summarizing results from an interview study we
conducted to inform the design of Foresight. 

} % end of abstract

\keywords{Exploratory data analysis, guided exploration, recommendation, 
statistics, visualization, sketching, scalable analytics.}

\usepackage{setspace}% http://ctan.org/pkg/setspace
\begin{document}



\firstsection{Introduction}
\maketitle
\label{sec:intro}
% !TEX root = paper.tex
% above command is for TeXShop

\begin{figure*}
  \centering
  \includegraphics[width=0.95\linewidth]{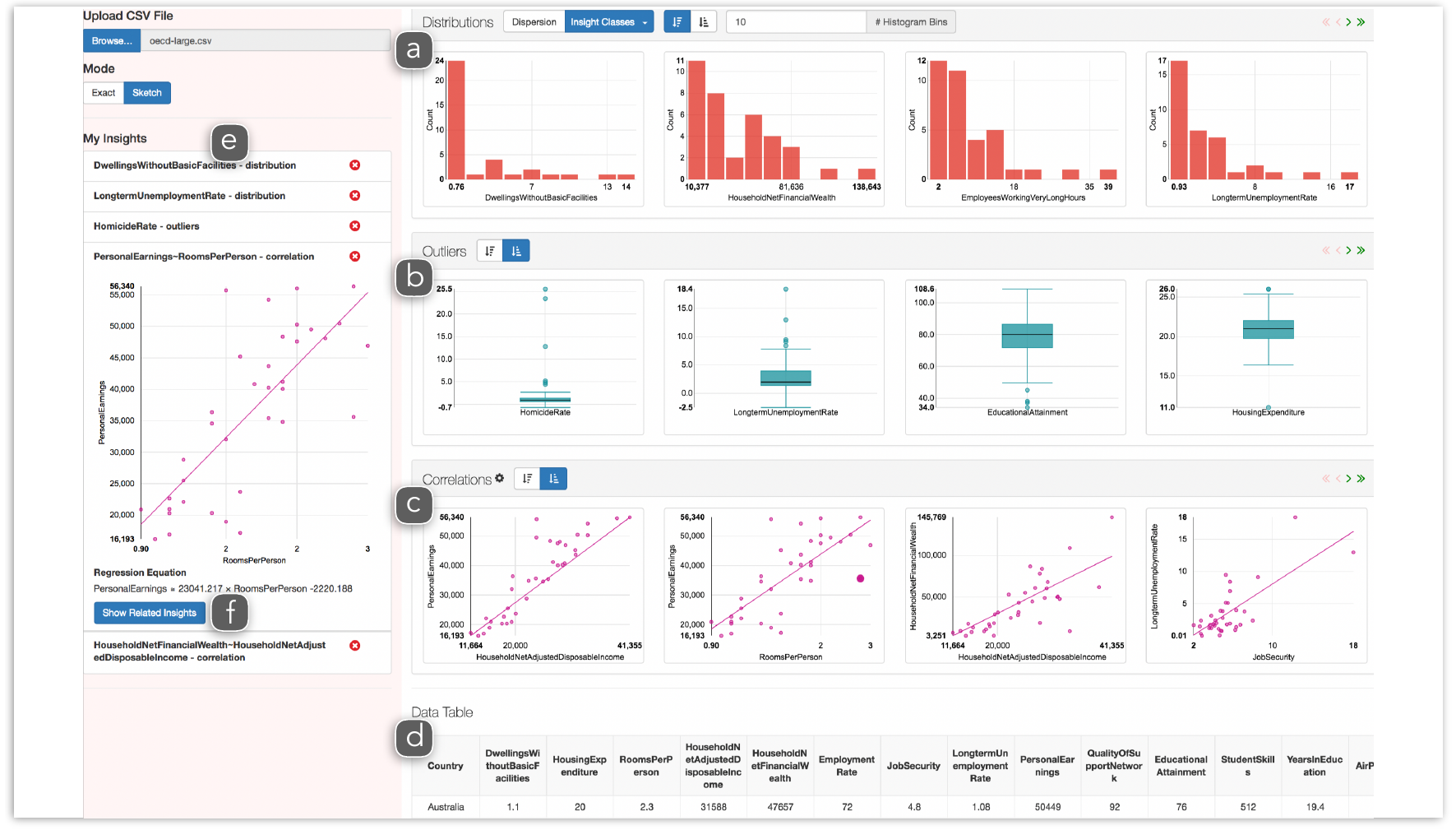}
  \caption{
    Each carousel in the Foresight user interface contains  a distinct
    collection of guideposts determined by a corresponding descriptor.
    Visualizations within a carousal are ranked by the descriptor's ranking
    metric with the strongest guidepost displayed first. The screenshot shows
    three descriptors (out of six supported by Foresight):  low dispersion (a),
    outliers (b), and linear correlations (c). Users can directly access the
    underlying data (d), bookmark (e) a guidepost  and explore ``related" (f)
    guideposts. Foresight enables fast response times for exploration of large
    high-dimensional data through approximate but provably good sketching
    techniques. \label{fig:insights}\label{fig:teaser} }
  \end{figure*}

%Data analysis is an iterative process with overlapping exploratory and
%confirmatory phases~\cite{tukey1985computer}; exploratory data analysis (EDA)
%is usually the first step.
Exploratory data analysis (EDA)~\cite{tukey1985computer} is a fundamental
process for understanding and reasoning about data in which analysts search for
interesting patterns and relations, ask questions, and (re)form hypotheses
explaining these patterns or relations. To this end, analysts derive insights
from the data by iteratively computing and visualizing summary statistics,
correlations, outliers, empirical distributions, density functions, clusters,
and so on.

\bfparhead{EDA Challenges:} Although the capabilities of EDA tools
continue to improve, most tools often require the user to manually select among
data attributes, decide which statistical computations to apply, and specify
mappings between visual encoding variables and either the raw data or the
computational summaries. This task can be daunting for large datasets having
millions of data items and hundreds or thousands of data attributes per item,
especially for typical users who have limited time and limited skills in
statistics and data visualization. Even experienced analysts face cognitive
barriers in this setting. 
%human working memory has inherent limited capacity and transient storage
%properties that limit the number of hypotheses, the amount of evidence, and
%the number of evidentiary relations that can be simultaneously attained,
Limitations on our working memory can cause large complex data to be
overwhelming regardless of expertise, and our tendency to fit evidence to
existing expectations and schemas of thought make it hard to explore insights
in an unbiased and rigorous manner~\cite{Pirolli_2005,Tversky_1975}. 
Thus, people typically fail both to focus on
the most pertinent evidence and to sufficiently attend to the disconfirmation
of hypotheses~\cite{nickerson1998confirmation}.
% select the evidence that is most diagnostic---i.e., most effective
% for choosing between given competing hypotheses---and fail to sufficiently
% attend to the disconfirmation of hypotheses~\cite{Pirolli_2005}.
Time pressures and data overload work against the analyst's ability to
rigorously follow effective methods for generating, managing, and evaluating
hypotheses.

\bfparhead{\thetool:} We address this problem by introducing \emph{\thetool}, a
system that facilitates rapid discovery of large, high-dimensional datasets
through ``guideposts.'' In \thetool, a guidepost is a visualization
corresponding to a pronounced instance of a statistical descriptor of the
underlying data. Examples of descriptors include dispersion, skew,
distributional tail weight, linear correlation between attributes $x$ and $y$,
clustering of $(x,y)$-values according to $z$-values, and the presence of
$x$-value outliers. A guidepost would then correspond to \emph{low} dispersion,
\emph{strong} linear relationship, \emph{tight} clustering, and so on.  \thetool
enables a user to jump-start the exploration process from automatically
recommended visualizations, and then gives the user increasing control over the
exploration process as familiarity with the data increases.  The resulting
efficiency in exploration  can save users time and dramatically improve their
productivity, thereby expanding the depth and breadth of generated hypotheses.
Our approach, in which visualizations  are recommended according to objective
criteria, also helps analysts focus their attention on evidence that is highly
diagnostic for, or disconfirming to, current hypotheses.

\bfparhead{Structured Exploration:} The key idea is to facilitate exploring data
attributes that have strong statistical features. This imposes a structure over
the space of data visualizations that leads to efficient navigation, rather
than mere ad hoc wandering.
%as in recent visualization recommenders, e.g.,
%\cite{Siddiqui_2016,Wongsuphasawat_2016}. 
We build on earlier work in guided and intelligent analytics systems, both
research and commercial~\cite{WatsonAnalytics,MSPowerBI,seo:infovis04,Wills_2008}. 
Each statistical descriptor in \thetool has one or more \emph{strength metrics} that
we use for ranking guideposts, e.g., the Pearson correlation coefficient to
measure the strength of a linear relationship. Similarly, each descriptor has
one or more visualizations, i.e., chart types, that we have selected by
considering established practice as well as findings of graphical perception
research. For a given descriptor, different metrics impose different rankings
on the descriptor instances; \thetool allows users to interactively change
ranking metrics to enable structured exploration of the data from multiple
perspectives.% (see Section~\ref{sec:model}). 

%Given an unfamiliar, complex dataset, the user can select one or more
%preliminary descriptors  to investigate. In this first, open-ended stage of
%exploration, for each descriptor, \thetool displays the strongest guideposts,
%i.e., visualizations of the specific sets of attributes having the highest
%values of the strength metric. In other words, the system displays to the user
%the result of a top-k ``guidepost query''. Using an iterative procedure, the
%user can dive deeper during a second level of exploration by examining
%recommended ``neighbors'' of a current guidepost or by taking control of the
%exploration process and adding to a guidepost query explicit constraints on,
%e.g., the data attributes considered or the values of the strength metric.

%Finally, \thetool  supports a third level of exploration by providing an
%\textit{overview} visualization for each descriptor to help orient the user and
%ensure that the exploration process is thorough.

\bfparhead{Scalability Through Sketching:} To achieve interactive 
performance when ranking guideposts, we use approximation techniques. 
Specifically, the dataset is preprocessed to compute sketches that will 
support fast approximate guidepost querying. 

%We demonstrate using experiments 
%on synthetic and real world datasets that, when presented with high dimensional
%data ($\geq 1000$ columns), Foresight's sketch based computation can
%consistently achieve over $90\%$ accuracy in statistical computation while
%consuming less than a third of the response time required by exact
%computation.
% 
% 
Overall, we contribute (1) an interview study providing insights into
EDA practices of data scientists, (2) a novel framework for exploring datasets
through ranked and neighborhood-based visualizations of data attributes that
manifest strong statistical properties,(3) an exploration engine supporting a
faceted interface for recommending and selecting visualizations that satisfy
user-specified constraints on data attributes and ranking-metric values, and
(4) sketch composition for fast approximate computation of ranking metrics and
visualizations.

In the following, we first provide a usage scenario to demonstrate how an
analyst might employ \thetool, followed by a discussion of prior work.  We then
present an interview study that we conducted with data scientists to inform our
design of \thetool and discuss the design considerations arising from both our
interview findings and prior art. %Next, we introduce the elements of \thetool,
%including that of the system design and the user interface, applying the design
%considerations that we developed. 
Next, we present details of our framework,
including the measures used for ranking and recommending guideposts and the
sketching algorithms applied for scalable computation of the statistical
descriptors. We conclude the paper by discussing future research directions.
%considerations in designing future visual analysis
%recommendation tools and 
%We follow with results from a performance evaluation, which demonstrate the
%benefits of sketch based computation in Foresight. Next, we present findings
%of a user study that we conducted to compare \thetool with a recent
%visualization recommender tool~\cite{Wongsuphasawat_2016}, suggesting that
%\thetool enables broader and faster exploratory data analysis.  

\section{Usage Example}
\label{sec:usageexample}
% !TEX root = paper.tex
% above command is for TeXShop

  We describe how an analyst uses \thetool to explore a dataset
  containing well-being indicators for the OECD member countries. This dataset
  contains 25 distinct attributes (indicators) for about 35 countries and is
  included in our demo mainly as an illustration and for ease of
  comprehension. \thetool is intended to facilitate interactive exploration
  of datasets with data items of the order of 100K having hundreds of 
  attributes.

  The analyst loads the OECD dataset in Foresight and eyeballs
  various guideposts displayed in the carousels corresponding to
  each descriptor type (Figure~\ref{fig:teaser}a-c).  She starts by 
  checking univariate distributions as ranked by \thetool. To gain 
  intuition on  how the system ranks the histogram distributions, she flips
  between ranking criteria, and decides that she is happy with the system's
  suggested ranking. She notices that \textsc{Personal Earnings} has the usual,
  skewed bimodal shape of an income distribution.  She can flip through the
  distributions carousel and bookmark (Figure~\ref{fig:teaser}e)
  interesting guideposts to revisit later. In particular she bookmarks the
  \textsc{Employees Working Very Long Hours} histogram, quickly checks the data
  points at the long tail of the distribution and takes note of the countries
  Turkey, Mexico, and Japan. 
  %The analyst wonders why Japan is there.

  She observes that a few data attributes have normal distributions but most
  have right or left skewed distributions.  She switches to the skewness
  criterion of \thetool ranking to focus on highly skewed attributes. \thetool
  suggests  \textsc{Job Security} has the highest skew. Our analyst notes
  Turkey and Mexico are among the countries with lowest job security and Japan
  is the country with highest job security.  She next examines the outlier
  guideposts provided by \thetool.  Enjoying the ease with which she can
  identify attributes having significant outliers, she saves the first three
  outlier guideposts, \textsc{Homicide Rate}, \textsc{Long Term Unemployment
  Rate}, and \textsc{Education Attainment}. She also notes that Turkey and
  Mexico are outliers in \textsc{Education Attainment}. Mexico is an outlier in
  \textsc{Homicide Rate} but Turkey is not. Neither Turkey and nor Mexico seems
  to suffer, however, from a high \textsc{Long Term Unemployment Rate}. Greece
  and Spain are outliers in this category.

  Already having a good sense of the individual data attributes, our analyst
  moves to examine the correlations recommended by \thetool. She flips the
  ranking order and notices that \textsc{Self Reported Health} has no
  correlation with \textsc{Time Devoted To Leisure}. Intrigued with this lack
  of correlation, she brings this guidepost \textit{into focus} by 
  bookmarking and clicks on the related guideposts button.  \thetool updates its recommendations by
  choosing a subset of guideposts within the neighborhood of the guidepost. The
  analyst explores the newly recommended correlations through multiple ranking
  metrics such as Pearson correlation coefficient, and quickly sees the
  guidepost showing a strong negative correlation between \textsc{Working Long
  Hours} and \textsc{Time Devoted To Leisure} since it is one of the top ranked
  correlation guideposts recommended. She finds this assuring but adjusts the
  correlation threshold slider to filter out very high correlations to focus on
  non-trivial relations at this stage, given the data attributes are
  particularly descriptive. 
  
  The analyst then goes back to the univariate
  distributional guideposts. The recommendations within these classes which
  have already been updated based on the previous selection show that
  \textsc{Time Devoted To Leisure} has a Normal distribution while \textsc{Self
  Reported Health} has a left-skewed distribution. Having gained greater
  familiarity with the OECD dataset, our analyst wonders about the factors that
  affect \textsc{Self Reported Health}. She clicks on the distribution of
  \textsc{Self Reported Health} adding this as one of the focal guideposts.
  Foresight recommends a new set of correlated attributes in which she finds
  that \textsc{Life Satisfaction} and \textsc{Self Reported Health} are highly
  correlated.

  %about \thetool showing  she finds with this quick discovery,
  %she brings this guidepost \textit{into focus} and clicks on the related guideposts button.
  %\thetool updates its recommendations by choosing a subset of guideposts within the neighborhood
  %of the focused guidepost. The analyst explores the newly recommended correlations
  %through multiple ranking metrics such as Pearson correlation coefficient,
  %and is surprised to learn that \textsc{Time Devoted To Leisure} has no correlation 
  %with \textsc{Self Reported Health}.
  Satisfied with her preliminary discoveries (and armed with deeper questions
  about OECD countries than ever before), our analyst saves the current
  \thetool state to revisit later.

\section{Related Work}
\label{sec:related}
% !TEX root = paper.tex
% above command is for TeXShop

\thetool builds on earlier work on automated visualization design,
visualization recommendation systems and scalable data analytics.
Prior work has proposed models and tools
(e.g., \cite{Mackinlay_1986,
roth:chi94,casner:tog91,Demiralp_2014a,avd2014infovis}) to automatically design
effective visualizations, building on Bertin's study~\cite{Bertin_1983} of
visual encoding variables and earlier graphical perception research, e.g.,
~\cite{cleveland:jasa84,shepard:sci87,lewandowsky:jasa89,
maceachren:book95,Mackinlay_1986,shortridge:ac82}. We consider established
practices in statistical graphics as well as findings of graphical perception 
research in deciding chart types and visual encoding for guideposts.    

Earlier work has introduced interactive 
systems and recommendation schemes~\cite{seo:infovis04,scagnostics2005infovis,showme:infovis07,Wills_2008,
Gotz_2008,Vartak_2015b,Vartak_2015a,Bouali_2016,Wongsuphasawat_2016,
Siddiqui_2016,Wongsuphasawat_2017,topk2017sigmod} to guide users in 
exploratory data analysis and visualization design.
%enhancing the breadth and depth of the
%exploration in data and design spaces. 
\thetool builds on this work and is
closest in underlying principles to the Rank-by-Feature
framework~\cite{seo:infovis04} and AutoVis~\cite{Wills_2008}. Both tools use
statistical criteria over data attributes in recommending or ranking
visualizations. The Rank-by-Feature framework also uses overview visualizations
of statistical features. \thetool differs from Rank-by-Feature
and AutoVis in three aspects. (1) \thetool employs a basic but larger set of
statistical descriptors frequently used by analysts in EDA and provides
browsable, faceted views with more flexible user control over ranking 
metrics.  (2) \thetool considers a notion of neighborhood, enabling users to effectively
explore visualizations related to an anchored (focused) visualization. (3) Also,
\thetool enables the fast approximate computation of statistical descriptors
through sketching. 

%\thetool is also related to earlier work on generating visualization sequences
%and complements this work in using statistical motivated transition costs.

Researchers have long realized that merely applying raw processing power by
itself is insufficient to guarantee interactive response
times~\cite{HellersteinACHORRH99}, and proposed methods that fall 
into one of the two categories in general: precomputation and sampling~\cite{hellerstein15}.  In
data visualization, \emph{precomputation} has traditionally referred to
processing data into formats such as prespecified tiles or cubes to
interactively answer queries via zooming, panning, brushing, and so on; see,
e.g., \cite{BattleCS16,kandel2012profiler,liu2013immens,lins2013nanocubes}. 
To efficiently compute statistical descriptors over large datasets, 
\thetool uses sketching~\cite{CormodeGHJ12}, 
an approximate querying technique based on the precomputation of 
\emph{synopses} of a dataset that can subsequently 
be composed to rapidly compute approximations to the exact 
query responses.

\section{Interview Study}
\label{sec:interview}
% !TEX root = paper.tex
% above command is for TeXShop

To inform the design of \thetool, we conducted semi-structured interviews. Our
goal was to obtain a preliminary understanding of EDA practices, patterns and
challenges. Our study follows~\cite{Kandel_2012} in part, but focuses on the
EDA aspects of data analysis in the context of predictive modeling.
We recruited ten participants from data science
teams within IBM Research by directly emailing them. 
%Through their work with IBM clients, they all had experience
%with datasets from diverse domains including healthcare, marketing, finance,
%retail, airline, social science, and simulation. Two of the participants were female and eight were male.
%Eight participants had PhD degrees and two had Master's degrees, all in science
%or engineering. 
%The most frequent data analysis goal among the participants was predictive
%modeling. 
Although our pool of participants may not be completely representative of the
broad community of data analysts, we found their responses to be very helpful
in designing \thetool. We aimed to find answers for the following questions in
our interviews.
\begin{itemize}[leftmargin=5pt,parsep=0pt,itemsep=0pt,topsep=2pt]
  \item[]How do analysts start exploratory data analysis?
  \item[]What tools do analysts generally work with?
  \item[]What visualizations and statistics do analysts frequently use?
  \item[]How do analysts decide on what is ``interesting'' in data?
  \item[]What strategies do analysts use with large data?
  \item[]What are productivity challenges in general and for specific tools?
\end{itemize}
%\parhead{Protocol:} We began each interview with a quick introduction describing
%the purpose of the interview. Each interview lasted about 30 minutes and was recorded
%with an audio recorder. During the interviews, three interviewers took notes independently.
\noindent We interviewed participants in person or via video conferencing.   
Each interview was recorded with an audio recorder while three interviewers 
took notes independently.  
%We analyzed our interview data using an iterative coding method. We merged the
%independent notes from the three interviewees, grouped common patterns and
%themes into categories. 
We now summarize our findings. 
\subsection{Results}
\bfparhead{EDA in Data Analysis Process:}
Where does EDA fall in the data analysis pipeline? 
%The overall analysis goal
% for the participants was predictive modeling. 
The interviews indicate that EDA
primarily happens between profiling and modeling tasks.
%(\autoref{fig:eda}).
%\begin{figure}[!h]
  %\centering
%\includegraphics[width=\linewidth]{figures/eda.pdf}
%\caption{Exploratory data analysis falls between profiling and modeling.\label{fig:eda}}
%\end{figure}
Once the data was ready for analysis, our participants (10/10) spent most of
their time trying to understand  the ``nature'' of the data.  Within this
process, analysts spent considerable time on first-order understanding: what
data attributes meant, how they were correlated,  how they related to each
other semantically and causally in the given data domain.

\bfparhead{Junior versus Senior Analysts:}
The interviews suggested a clear separation between senior and junior data
analysts in how they approach EDA.  Senior analysts, defined as having more
than five years of experience, spent a significant amount of time on domain
understanding through close collaboration with clients.  They emphasized the
importance of understanding semantic relationships and relied less on machine learning based automated techniques than junior analysts. Junior analysts transitioned faster
from the EDA phase to the modeling phase than did senior analysts. 
%Junior
%analysts also used more complex statistical techniques than senior analysts
%when deciding what was interesting or important in data. Senior analysts relied
%on basic statistical techniques while putting more emphasis on the importance
%of understanding the domain-specific causal relationships underlying the data.

\bfparhead{Stratified Greedy Navigation:}
Results indicate that analysts use a stratified navigation strategy in their
analysis, moving from simpler, univariate properties to more complex,
multivariate relations. They started the exploration by examining  attribute
names and what they might mean, and then computed quantities such as min, max,
summary statistics, and, for categorical variables, the most frequent data
values. They then moved to univariate densities and histograms, and looked for
outliers. Analysts used basic visualizations related to these measures,
including bar charts of histogram density estimates, box plots, and Pareto
charts.

Next, they looked at bivariate relations, primarily computing correlations and
visualizing them with scatter plots. Only one analyst considered trivariate
correlations. 
%Computations of higher-order relations or patterns varied across analysts and
%depended on the particularities of the data, domain, and predictive modeling
%task at hand: Linear regression (5/10), logistic regression (5/10), lasso-based
%feature selection (2/10), hierarchical clustering (4/10), and dimensionality
%reduction (4/10). 
Such a stratified, low-to-high order work flow allows 
analysts to terminate the EDA phase as soon as they are satisfied
with the results, thereby minimizing effort~\cite{Pirolli_2005}. 
%These 
%findings are broadly inline with the EDA strategy, \emph{examine each dimension first and then find relationships among dimensions}, proposed \etal~\cite{Huberty_1991}.  

As analysts moved up in the exploration hierarchy, they used a greedy strategy
for focusing on the attributes. 
%When asked about how they ensured that they
%didn't miss relevant patterns or relations in the broader data space, junior
%analysts (2) mentioned using ML based automated techniques. Two senior
%analysts, however, noted the lack of better alternatives.  
We conjecture that the greedy strategy might also be motivated by a desire to
minimize cognitive costs. This strategy is dangerous, though: an analyst can
become ``trapped'' at a local ``optimum'' and thereby miss important, relevant
insights.

\bfparhead{Tools:}  Our analysts had multiple tools in their toolset, aimed at different purposes. 
All participants used Python together with one or more Python libraries/tools.  
% including
%Pandas, Jupyter, scikit-learn, Keras, PySpark, and the Python TensorFlow wrapper. 
Those participants who had used primarily R, Matlab or Java (Weka) in the
recent past were transitioning to greater use of Python, because it supports a
broad range of data analysis tasks in a scalable manner. 
%The shift can be also attributed to the company culture as many data science
%teams in IBM use Python as their primary tool.  
R was used for its rich statistical computation and particularly for its
popular visualization library (ggplot).
Participants frequently used visualizations such as histograms, scatter plots,
box plots and q-q plots.  Some also noted using
dimensionality reduction  as well as clustering, visualized via scatter plots,
dendogram trees and heat maps.  
%One analyst asserted that ``I don't really care
%about what visualization I use or whether it looks good or not. I care about
%insight.''  She followed with an example of a high level insight: {\em Beer
  %sales increase 5\% with 5 degrees decrease in weather temperature}  and noted
  %``give me that!''
%\begin{figure}[!h]
  %\centering
  %\includegraphics[width=\linewidth]{figures/eda-tools}
%\caption{Tools and visualizations used by study participants.\label{fig:eda-tools}}
%\end{figure}

\bfparhead{Handling Big Data:}
Most of the analysts (8/10) had experience with big-data analysis.  They
primarily used random and guided sampling. Some (2/8) applied techniques to
match the sample distribution to that of the original data. Analysts performed
their exploratory data analysis on sampled datasets and applied predictive
modeling to the complete data.  In general, our analysts were not concerned
with offline scalability for predictive modeling and noted that they
were satisfied with capabilities of, e.g., the Python libraries such as Pandas
and PySpark. The main problem was that the prediction computations had to be
done outside the exploratory data analysis workflow.

\bfparhead{Challenges:}
%Participants (10/10) regularly worked with clients. Data sets were given to
%them in a state ready for analysis in many cases.  In other cases, however,
%they needed to wrangle and profile the data themselves. Participants indicated
%that cleaning the data, getting it into the right format, and dealing with
%missing values or measurements were the most painful aspects of their workflow.
%These issues have been widely reported in prior work; see, e.g.,
%\cite{Kandel_2012}.
Once the data was readied for analysis, the time spent on EDA dominated the
time spent on either modeling or reporting. The main productivity challenge for
most of our participants (7/10)  was not knowing either where to start or how
to go about effectively exploring the data attributes. The problem was
exacerbated when the data had many attributes, as participants needed to
explore a large number of both individual features and correlated feature pairs
when creating a model. New analysts can easily feel overwhelmed.  One junior
analyst noted that  complexity of the tools (``many options and different ways
of achieving the same functionality'') magnified the required amount of time
and effort.
% Mapping analytical to intend to a program

\section{Design Criteria}
\label{sec:criteria}
% !TEX root = paper.tex
% above command is for TeXShop

Our design of \thetool is influenced by the following criteria.  We determined
these criteria based on our findings from the interview study, our own experience
in EDA and prior research.

\parhead{1.{\enspace}Organize exploration around statistical descriptors.} The system should treat descriptors as first class entities and organize its visualizations and recommendations around
them.  \thetool supports a wide variety of descriptors, which in turn determine the data dimensions and visualizations that are shown.

\parhead{2.{\enspace}Use descriptor strength to drive the promotion of data variation.}
The system should enable the exploration of the data attributes based on the
interesting statistical features that they possess. \thetool shows
carousels corresponding to a range of different descriptors, and ranks specific instances according to an insight-specific metric within each carousel. 

\parhead{3.{\enspace}Give user control over the definition of descriptor strength.} Users
should be able to change the metric used to rank the instances of a descriptor.  \thetool
enables users to interactively change the ranking metrics.

\parhead{4.{\enspace}Enable scalable analytics.} The system should enable
exploratory data analysis with large data. \thetool uses sketching
algorithms to effectively compute approximate rankings and overview visualizations over large data sets, and allows users to
override system defaults for sketch-based computations.

\parhead{5.{\enspace}Use the best visualizations for communicating statistical descriptors.} The tool should display its recommendations using the most appropriate visualizations for each descriptor. When constructing guideposts, \thetool adapts
visualizations informed by best practices and graphical perception.

\parhead{6.{\enspace}Facilitate stratified work flow to minimize the cost of exploration.} The system should support focused as well as broad exploration in a layered, simple-to-complex fashion. In \thetool, the carousels are ordered so that univariate descriptors are shown first. Each carousel provides a different perspective into properties of the data.  Crucially,
\thetool also supports focused exploration and recommends ``related'' guideposts anchored at the focused guidepost.

\parhead{7.{\enspace}Enable access to raw data on demand.} Users should able
to directly  examine the raw data, without leaving the context of data exploration.
\thetool displays the data in a table view and enables fluid transitions
between guideposts and the data rows and columns governing the guideposts.

%\section{User Interface and System Design}
%\label{sec:ui}
%\input{interface}

\section{Guideposts}
\label{sec:model}
% !TEX root = paper.tex
% above command is for TeXShop

In this section, we first describe the basic concepts of guideposts and guidepost queries, and the statistical descriptors on which they are based. We then outline the set of descriptors currently supported in our \thetool prototype.

\subsection{Exploring Data with Guidepost Queries}

The input data to \thetool is a matrix $A_{n \times d}$, where each row
represents one of $n$ \emph{data items} and each column represents one of the
$d$ \emph{attributes} of an item. A \emph{descriptor} of the data is a
statistical property defined in terms of $m$ data attributes for some
$m\in\{1,2,\ldots,d\}$; we focus throughout on descriptors with $m=1$, such as
the dispersion or skew of an attribute, or $m=2$, such as linear correlation or
two-dimensional clustering for a pair of attributes. The \emph{instance set}
for a descriptor is the set of $m$-tuples of attributes upon which the
descriptor can be defined. For example, given a data set with attributes
$a_1,a_2,\ldots,a_d$, the instance set corresponding to the descriptor
$\mathcal{I}=$ ``linear correlation" would contain all pairs $(a_i,a_j)$ with
$i<j$ such that $a_i$ and $a_j$ are both real-valued attributes. We require
that each descriptor have one or more associated \emph{strength metrics} that
can be used to rank the $m$-tuples of attributes in the instance set, e.g., the
Pearson correlation coefficient for the descriptor~$\mathcal{I}$ above.
Similarly, each descriptor must have one or more associated \emph{data
visualizations}, i.e., charts. 

A \emph{guidepost} for a given descriptor and strength metric is a data
visualization corresponding to an $m$-tuple of attributes from the descriptor's
instance set, where the member is highly ranked according to the strength
metric. A descriptor can also have one or more associated \emph{overview
visualizations} that display the values of a strength metric over all tuples in
the instance set. For example, an overview visualization for the
descriptor~$\mathcal{I}$ above might comprise a heat map where the $x$ and $y$
coordinates correspond to the different attribute pairs and the color and the
size of a circle centered at $(x,y)$ encodes the Pearson correlation
coefficient.

A basic \emph{guidepost query} for a descriptor returns the visualizations for
the highest-ranked attribute tuples according to the selected descriptor
strength metric, e.g., the attribute pairs with the highest Pearson
correlations. This represents the first level of exploration, where the user
selects a descriptor and the system returns the strongest examples of its
instance set.

In a second level of exploration, the user can gradually take control of the
exploration process as they become familiar with the data. In particular, the
system can recommend guideposts that are ``nearby" to a particular guidepost
selected by the user.  As a simple example, suppose that the initial
descriptor is $\mathcal{I}=$ ``linear correlation" as above. The system
executes a basic guidepost query and produces visualizations corresponding 
to the $(x,y)$ pairs having the highest Pearson correlation coefficient. 
Suppose the user brings into focus a particular guidepost, i.e., the
visualization for a particular pair $(\bar{x},\bar{y})$. One natural notion of
a ``neighborhood" $N_{\bar{x}}$ of $(\bar{x},\bar{y})$ can be defined as the
set of pairs of the form $(\bar{x},y)$ having the highest correlation
coefficients. That is, we have fixed the $x$ attribute and only allowed the $y$
attribute to vary in our search, thereby filtering the set of attribute pairs
considered. We can define a neighborhood $N_{\bar{y}}$ in a similar manner, by
fixing $y$, and can also define $N_{\bar{x},\bar{y}}$ as the set of most highly
correlated attribute pairs in $N_{\bar{x}}\cup N_{\bar{y}}$. 
%The user can
%assert even more control by formulating guidepost queries that explicitly
%specify constraints on the subset of attributes are to be considered.
%We focus on the foregoing notion of neighborhood in our initial prototype. As
%part of our ongoing research, we are identifying other useful notions of
%guidepost neighborhoods. Our system is designed to allow easy incorporation of
%new notions of ``neighborhood." For instance,  one can envision more
%elaborate notions in which, say, the neighborhood of a guidepost for single
%attribute $\bar{x}$ is defined in terms of the most highly correlated pairs of
%the form $(\bar{x},y)$.

%The user can also exert control by adding constraints or filters on the
%strength metric to their queries. For example, the user might want to
%rank only correlated attribute pairs whose correlation coefficient falls in the
%range $[0.5,0.8]$ because they want to filter out extremely high
%correlations---such correlations can correspond to trivial associations, as
%with a pair attributes that measure the same quantity, one in pounds and the
%other in kilograms.

%Constraints with respect to statistical significance, in the spirit of, e.g.,  \cite{ZhaoSZBUK17}, can be incorporated directly into the strength metric. For example, a modified metric value for linear correlation can be obtained by multiplying the Pearson correlation by zero if the $p$-value of an associated test statistic is less than an appropriate significance value. In ongoing work, we are exploring other ways of incorporating metrics of statistical significance metrics into \thetool.

In a third level of exploration, at any point during the EDA process the user
can step back and look at the overview visualization of a descriptor. This helps
ensure that, in the analogy with gradient descent, the EDA process does not get
inadvertently ``trapped" in some local ``neighborhood" of attribute tuples.
This capability is particularly important in cases where many attribute tuples
have similarly high descriptor-metric scores, so that the particular set
visualized for the user is somewhat arbitrary. In practice, the ability to
effectively navigate out of a local guidepost neighborhood helps analysts avoid
prematurely fixating on a particular set of data attributes
during exploratory data analysis.

\subsection{\thetool's Statistical Descriptors}

Foresight is designed to be an extensible system where new descriptors can 
be ``plugged in,'' along with their corresponding ranking measures and visualizations.  In
this section, we discuss the specific descriptors supported by our current
\thetool prototype. Denote by $\mathcal{B}$ and $\mathcal{C}$ the sets of
attribute columns in $A$ that contain numeric and categorical values. \thetool
currently supports six distinct descriptors of statistical properties, each
with a preferred ranking metric and visualization method. In our description,
we denote by $b=(b_1,\ldots,b_n)^\top\in\mathcal{B}$ a numeric column with mean
$\mu_b$ and standard deviation $\sigma_b$, and by
$c=(c_1,\ldots,c_n)^\top\in\mathcal{C}$ a categorical column. For each
descriptor, the \textit{ranking metric} is italicized.

\bfparhead{1.{\enspace}Dispersion:} Dispersion measures the extent to which the data is concentrated around the mean.
We use a robust scale-invariant statistical measure to quantify dispersion, namely the \textit{quartile coefficient of dispersion}. For a
given numeric column $b\in\mathcal{B}$, let $Q_1=Q_1(b)$ and $Q_3=Q_3(b)$ denote the first and third quartile values, i.e., the 0.25 and 0.75 quantiles. The quartile coefficient of dispersion is defined as $\text{qcd}(b) = (Q_3 - Q_1)/(Q_3 + Q_1)$. We visualize dispersion via a histogram.

%PJH: Can also be measured robustly by maximum absolute deviation around the median (MADAM)

\bfparhead{2.{\enspace}Skew:} Skewness is a measure of asymmetry in a univariate distribution.
We measure skewness with \textit{standardized skewness coefficient}
$\gamma_1(b) = n^{-1}\sum_i^n (b_i - \mu_b)^3/\sigma_b^3$
%$\gamma_1(b) = \frac{n^{-1}\sum_i^n (b_i - \mu_b)^3}{\sigma_b^3}$
and visualize it via a histogram.

\bfparhead{3.{\enspace}Heavy tails:} Heavy-tailedness is the propensity of a distribution towards extreme values.
We measure heavy-tailedness with \textit{kurtosis}
$\text{Kurt}(b) = n^{-1} \sum_i^n (b_i - \mu_b)^4/\sigma_b^4$
%$\text{Kurt}(b) = \frac{n^{-1} \sum_i^n (b_i - \mu_b)^4}{\sigma_b^4}$
and visualize it via a histogram.

\bfparhead{4.{\enspace}Outliers:} We measure the presence and significance of outliers or extreme values using the Tukey test. In a Tukey box-and-whisker plot, the ``inlier range" between the whisker endpoints for a column $b=(b_1,\ldots,b_n)^\top$ is defined as $[Q_1 - 1.5 (Q_3 - Q_1), Q_3 + 1.5 (Q_3 - Q_1)]$, where $Q_1$ and $Q_3$ are the first and third quartile values as before. Data values $b_i$ that fall outside this range are considered outliers. We visualize outliers using box-and-whisker plots, and use the \textit{number of outliers} as the strength for ranking. 
%Note that one can envision alternative ranking metrics that consider both the number and extremity of outlier points---we focus on number of outliers for simplicity.
% \noindent\textbf{5. Multimodality:} For a numeric column $b$, the \textit{number of modes} is the number of local peaks, suitably defined, in the (estimated) probability density function for continuous data or in the probability mass function for discrete data. For a categorical column $c$, the \textit{number of modes} is simply the number of elements with maximal frequency of occurrence. Multimodality is visualized via a histogram.

\bfparhead{5.{\enspace}Heterogeneous frequencies:} For a categorical column $c$ (or a
discrete numerical column $b$), high heterogeneity in frequencies implies that
a few values (``heavy hitters'') are highly frequent while others are not. We
measure the heterogeneity strength of column
$c=(c_1,\ldots,c_n)^\top\in\mathcal{C}$ using the \textit{normalized Shannon
entropy}. 
%Specifically, denote by $\mathcal{D}=\mathcal{D}(c)$ the set of
%distinct values appearing in $c$, by $D=|\mathcal{D}|$ the numer of distinct
%values, and by $n_d$ the multiplicity in $c$ of value $d\in \mathcal{D}$, and
%set $f_d=n_d/D$. Then the normalized Shannon entropy is defined as  $E(c) =
%\sum_{d\in \mathcal{D}} f_d\log{f_d}/\log{D}$. (The normalizing constant
%$\log{D}$ ensures that $0\le E(c)\le 1$.) 
We visualize heterogeneity in frequencies with a Pareto chart.
%PJH: Could also use coef. of variation of frequencies as a measure.

\bfparhead{6.{\enspace}Linear Relationship:} We measure the strength of a linear relationship between two columns
$x, y \in \mathcal{B}$ using the \textit{absolute value of the Pearson correlation coefficient}
$\rho(x, y) = \vert \sum_{i=1}^n (x_i-\mu_x)(y_i-\mu_y)/(\sigma_x\sigma_y) \vert$ and visualize it via a scatter plot with the best-fit line superimposed.

\pagebreak
\section{Facilitating Scalability Through Sketching}
\label{sec:sketch}
% !TEX root = paper.tex
% above command is for TeXShop

%\begin{figure}[t]
  %{\centering
    %%\includegraphics[width=0.5\textwidth]{figs/foresightarchcropped}
    %\includegraphics[width=0.5\textwidth]{figs/foresightarch-shadowed}
    %\caption{\small Architecture of Foresight prototype. The Sketch \& Insight microservices are implemented as
      %RESTful APIs in the R language. The UI is implemented as an Angular-Meteor web app.\label{fig:foresightarch}}
    %}
  %\end{figure}

We use sketching~\cite{CormodeGHJ12} to speed up the computation of strength 
metrics.  Some strength metrics are fast and easy to compute. For instance, skewness and
kurtosis can both be computed for numeric columns in a single pass by
maintaining and combining a few running sums. For the remaining metrics,
sketches---lossy compressed representations of the data---are crucial in order
to preprocess the data in a reasonable amount of time. \thetool integrates and 
composes a variety of sketching and sampling techniques from the
literature, namely quantile sketch, entropy sketch, frequent items sketch,
random hyperplane sketch, and random projection sketch; see, e.g.,
\cite{CormodeGHJ12}. 
For example, we precompute a sketch $b^*=\phi(b)$ for each numeric $n$-row column $b\in\mathcal{B}$, where $\phi$ is a suitable randomly chosen mapping to a binary vector of length $k\ll n$. Then a good approximation to the correlation coefficient between columns $x$ and $y$ can be quickly computed on the fly as $\rho(x,y)\approx \cos(\pi H_{x^*y^*}/k)$, where $H_{x^*y^*}$ is the number of entries where $x^*$ and $y^*$ differ.
Our initial experiments (without parallelism) 
show $>90\%$ accuracy and $3x-4x$ speedup in preprocessing.
%,
%with interactive speeds during exploration.
Due to space restrictions, we defer the complete details of
sketch composition to the full research version of this paper.

\section{Conclusion and Future Work}
\label{sec:conclusion}
% !TEX root = paper.tex
% above command is for TeXShop

We present \thetool, an interactive system that enables rapid, structured
exploratory data analysis via statistical guideposts. Our approach uses
recommendations to guide users in exploring unfamiliar large and complex
datasets, and gradually gives them more and more control over the exploration
process. We report insights from an interview study into EDA practices 
of data scientists, informing the design of EDA tools at large.

We have introduced a basic notion of neighborhood search and will
investigate the topology of visual analysis space more deeply in future work. 
%As part of our ongoing research, we are identifying other useful notions of
%guidepost neighborhoods. 
Our system is designed to allow easy incorporation of different notions of
``neighborhood." For instance,  one can envision a notion in
which, say, the neighborhood of a guidepost for single attribute $\bar{x}$ is
defined in terms of the most highly correlated pairs of the form $(\bar{x},y)$.

\thetool enables users to exert control over guidepost rankings by adding
constraints or filters on the strength metric to their queries. Constraints
with respect to statistical significance, in the spirit of, e.g.,
\cite{ZhaoSZBUK17}, can be incorporated directly into the strength metric. For
example, a modified metric value for linear correlation can be obtained by
multiplying the Pearson correlation by zero if the $p$-value of an associated
test statistic is less than an appropriate significance value.

Using sketching and indexing methods, our system
can currently handle datasets with large numbers of rows and moderate numbers
of columns. We plan to improve scalability with respect to the number of columns by
incorporating parallel search methods that speed up guidepost queries. We also aim to support additional descriptors, e.g., scagnostics as in \cite{scagnostics2005infovis}.

Finally, in ongoing work, we are evaluating  \thetool through a human-subjects
experiment with data scientists to better understand the merits and
limitations of our approach. 

% For example, the user might want to
%rank only correlated attribute pairs whose correlation coefficient falls in the
%range $[0.5,0.8]$ because they want to filter out extremely high
%correlations---such correlations can correspond to trivial associations, as
%with a pair attributes that measure the same quantity, one in pounds and the
%other in kilograms.

%We introduce a novel approach to visualization recommendation via the notion of
%insights. Our approach uses recommendations to guide users in exploring
%unfamiliar large and complexdatasets, and gradually gives them more and more
%control over the exploration process.  We introduce a basic notion of
%neighborhood search and will investigate the topology of insight space more
%deeply in future work. We will also explore the practical efficacy of more
%expensive but more robust strength measures such as maximumabsolute deviation
%from the mean (MADAM). Using sketching and indexing methods, our demo system
%can currently handle datasets with largenumbers of rows and moderate numbers of
%columns.  Future work will improve the scalability with respect to columns by
%incorporating parallel search methods that speed up insight queries.

\balance{}

% REFERENCES FORMAT
% References must be the same font size as other body text.
\bibliographystyle{abbrv-doi}
\bibliography{paper}
\end{document}